\documentclass[aps,a4paper,english,nofootinbib,showpacs,twocolumn]{revtex4}
\usepackage{color}
\usepackage{amssymb}
\usepackage{amsmath}
\usepackage{graphicx}
\usepackage[T1]{fontenc}
\usepackage{hyperref}
\usepackage[brazil]{babel}
\usepackage[latin1]{inputenc}
\usepackage{amsfonts}

\newcommand{\be}{\begin{equation}}
\newcommand{\ee}{\end{equation}}

\newcommand{\ba}{\begin{eqnarray}}
\newcommand{\ea}{\end{eqnarray}}

\begin{document}

\title{A model to localize gauge and tensor fields on thick branes}
\author{A. E. R. Chumbes}
\email{arueda@feg.unesp.br}
\author{J. M. Hoff da Silva}
\email{hoff@feg.unesp.br, hoff@ift.unesp.br}
\author{M. B. Hott}
\email{marcelo.hott@pq.cnpq.br}
\affiliation{UNESP Universidade Estadual Paulista - Campus de Guaratinguet\'a - DFQ.\\
Av. Dr. Ariberto Pereira da Cunha, 333 CEP 12516-410, Guaratinguet\'a-SP,
Brasil.}
\pacs{11.27.+d, 98.80 Cq}

\begin{abstract}
It is shown that the introduction of a suitable function in the
higher-dimensional gauge field action may be used in order to achieve gauge
bosons localization on a thick brane. The model is constructed upon
analogies to the effective coupling of neutral scalar field to
electromagnetic field and to the Friedberg-Lee model for hadrons. After that
we move forward studying the localization of the Kalb-Ramond field via this
procedure.
\end{abstract}

\maketitle

\section{Introduction}

It is well known that in the braneworld paradigm four-dimensional gravity
may be localized on a singular brane \cite{RS}, i. e., a normalizable zero
mode arising from the gravitational field fluctuation exists on the brane.
In Ref. \cite{MG}, a non-singular brane performed by a thick domain wall is
considered. In this more realistic case, gravity is also localized on the
brane. In general, braneworld models are inspired in string theory and it is
expected that a considered model makes contact with some string theory limit
in the consideration, for instance, of $D$-branes solutions. In this vein,
it is important to make effort in order to eliminate some of the differences
between the domain-wall approach for braneworld models and the $D $-branes
solutions.

As already noted, for example, in Ref. \cite{KT}, an important difference of
$D$-branes when compared to the domain wall is that while the former
supports gauge fields living on it (basically arising from the open strings
ending on the $D$-branes), it is not always possible to achieve gauge fields
localization on the domain walls by means of only the spacetime curvature
acting, i. e., the gauge field effective action term is blind with respect
to the warp factor. In other words, as well known \cite{DGS}, the
five-dimensional gauge field action
\begin{equation}
-\frac{1}{4}\int d^{5}x \sqrt{g}F^{MN}F_{MN},  \label{1}
\end{equation}
simple blows up after the dimensional reduction. The simplest approach to
reach zero modes of gauge fields on the brane is by assuming the existence
of bulk gauge fields which could, in principle, to give rise to the
four-dimensional gauge sector on the domain wall. Unfortunately such an
approach indicate that the gauge fields cannot be localized \cite%
{DGS,DSI,DSII}.

In order to circumvent this difficulty, some models have appeared in the
literature. In the absence of gravity, gauge field localization was
extensively considered in, for instance, Ref. \cite{RUBA} (see also
references therein). In the context of curved spaces this issue was also
analyzed. In Ref. \cite{KT}, an additional scalar field --- the dilaton ---
introduced in the five-dimensional action is responsible to drive the gauge
field localization, by means of the coupling between the dilaton and the
kinetic term of gauge fields. Similar procedure was also adopted in \cite%
{CTA}. In Ref. \cite{melfo} gauge filed localization obtained via kinetic
terms induced by localized fermions. After all, however, it is relevant to
note that there is not a complete mechanism concerning gauge field
localization on the brane.

In this paper we shall add one more possibility in order to localize gauge
fields in thick branes. From the pragmatic point of view the idea is quite
simple and it is based on the same mechanism which provides the localization
of spin 1/2 fermion fields in a brane in five-dimensional flat \cite{RUBA1}
and warped \cite{GABA} space-time. We just introduce a suitable function in
the five-dimensional gauge field Lagrangian, which leads to a normalizable
zero mode after the dimensional reduction, namely:
\begin{equation}
-\frac{1}{4}\int d^{5}x\sqrt{g}G(\phi )F^{MN}F_{MN}.  \label{2}
\end{equation}%
The $G(\phi )$ is a functional of the scalar field from which the brane
originates. To fix ideas one should think in the model obtained in \cite{MG}%
. Therefore two questions are immediately raised: firstly, since the
inserted function depends on the scalar field, it should enter in its field
equations contributing to the background constituted by the metric and the
scalar field; secondly, how to set up the form of such a functional, since
any normalizable function could, in principle, act in the same way in the
gauge localization scheme. To the first point we should assume that in this
effective model $G(\phi )$ is a function of the minimum energy solution, $%
\bar{\phi}(r)$, which represents the brane (the domain wall solution), such
that there is no contribution of the gauge field zero-mode to the energy of
the system, as it happens in the localization of fermion zero-mode in the
brane. The second question is a little more subtle. While it is true that
the procedure explained in the next Section may be successfully repeated
with any normalizable function $G(\bar{\phi}(r))$, we shall give a physical
motivation based upon analogies to the Schwinger's neutral scalar-gauge
field coupling \cite{SCHWINGER}, to the color dielectric model for the
confinement of gluons and quarks \cite{LEE} and to the quantum mechanics
associated to the matter fields localization on branes.

Apart of that, it is also known that string theory presents plenty of higher
spin fields on its spectrum. Therefore it is quite conceivable the study of
such fields in the braneworld context. In this vein it is important to
analyze the possibility of localize the Kalb-Ramond (K-R) field \cite{KR} on
the brane. In the context of infinitely thin branes the localized zero mode
of the K-R field (interpreted as torsion) is highly suppressed by the size
of the extra dimension \cite{KRRS}. Within the framework of thick
braneworlds the K-R field was also investigated in \cite{TCA} and in \cite%
{CCT}. In fact, in \cite{TCA} it was demonstrated that there is no localized
tensorial zero mode with the usual thick brane background. It was shown
that, in order to localize the zero mode it is necessary a background
composed by a membrane described by two real scalar fields with internal
structures, or a dilatonic gravitation. Part of this paper is devoted to the
use of appropriated smearing out functions in order to localize the K-R zero
mode field on the brane. The aforesaid functions, as mentioned, are suitable
constructed based upon the same mechanism for the localization of spin $1/2$
fermion fields on a 3-brane embedded in flat \cite{RUBA1} and warped \cite%
{GABA} background.

The paper is structured as follows: in the next Section we show, in very
simple grounds, that the introduction of the $G(\bar{\phi})$ function do
localize normalizable zero-mode gauge fields on the brane, physically
motivating the functional form of $G(\bar{\phi})$. In Sec. III we take
forward our analogy applying, then, a similar procedure to localize the zero
mode of the K-R field and stressing an important point concerning the
integrability of the smearing out functions. In the last Section we conclude.

\section{Localizing Gauge Fields}

Before starting our analysis properly, let us briefly set the background by
recalling the standard model developed in Ref. \cite{MG} for
five-dimensional gravity coupled to a real scalar field:%
\begin{equation}
S=\int d^{5}x\sqrt{g}\Big(-\frac{1}{4}R+\frac{1}{2}(\partial \phi
)^{2}-V(\phi )\Big),  \label{3}
\end{equation}%
where the Poincare invariant line element is given by%
\begin{equation}
ds^{2}=e^{2A(r)}\Big(dt^{2}-\sum_{i=1}^{3}dx_{i}^{2}\Big)-dr^{2}.  \label{4}
\end{equation}%
By admitting that the scalar field is dependent on the extra dimension only,
the Einstein-Hilbert and scalar field equations admit minimum energy
solutions which are also solutions of the first-order differential equations
\cite{MG}
\begin{equation}
\frac{d\phi }{dr}=\frac{\partial W(\phi )}{\partial \phi }=W_{\phi }
\label{5}
\end{equation}%
and%
\begin{equation}
\frac{dA}{dr}=-\frac{2}{3}W(\phi ),  \label{6}
\end{equation}%
whenever the potential $V(\phi )$ is written in terms of the superpotential $%
W(\phi )$ as \cite{UM,DOIS}%
\begin{equation}
V(\phi )=\frac{1}{2}W_{\phi }^{2}-\frac{4}{3}W(\phi )^{2}.  \label{6a}
\end{equation}%
In \cite{MG} the superpotential is chosen to be given as%
\begin{equation}
W(\phi )=3bc\sin \Big(\sqrt{\frac{2}{3b}}\phi \Big),  \label{7}
\end{equation}%
which leads to%
\begin{equation}
A(r)=-b\ln (2\cosh (2cr))  \label{8}
\end{equation}%
and%
\begin{equation}
\bar{\phi}(r)=\sqrt{6b}\arctan \Big(\tanh \Big(cr\Big)\Big).  \label{9}
\end{equation}%
The free parameters $b$ and $c$ in this model are related to the thickness
of the brane $(c)$ and the AdS curvature $(bc)$.

Having fixed the background, let us study the standard protocol for gauge
field localization, this time armed with the smearing out $G(\bar{\phi})$
function. The gauge field Lagrangian is given by Eq. (\ref{2}). As remarked
before, we shall neglect the $G(\bar{\phi})$ contribution to the background
in this effective model. The field equations reads%
\begin{equation}
\partial _{C}\Big(e^{4A}G(\bar{\phi})g^{CE}g^{DB}F_{EB}\Big)=0.  \label{10}
\end{equation}

In the gauge $\partial _{\rho }A^{\rho }=0$ and $A_{4}=0$, decomposing the
field as $A_{\mu }=\sum_{n}a_{\mu }(x)\alpha _{n}(r)$ one arrives at%
\begin{equation}
m_{n}^{2}\alpha _{n}(r)+e^{2A}\Big[\alpha _{n}^{\prime \prime }(r)+\Big(%
\frac{G^{\prime }(\bar{\phi})}{G(\bar{\phi})}+2A^{\prime }\Big)\alpha
_{n}^{\prime }(r)\Big]=0,  \label{11}
\end{equation}%
where prime means derivative with respect to $r$. In order to set a typical
quantum mechanical problem, let us make the following transformation%
\begin{equation}
\alpha _{n}(r)=e^{-\gamma (r)}g_{n}(r).  \label{12}
\end{equation}%
By means of the identification $2\gamma ^{\prime }=2A^{\prime }+G^{\prime
}/G $ the first derivative term disappear and the result is the Schrödinger
equation:%
\begin{equation}
-g_{n}^{\prime \prime }(r)+\big[\gamma ^{\prime \prime }+(\gamma ^{\prime
})^{2}-m_{n}^{2}e^{-2A}\big]g_{n}(r)=0.  \label{13}
\end{equation}%
For the massless zero mode $(g_{0}\equiv g)$ we have simply%
\begin{equation}
-g^{\prime \prime }(r)+[\gamma ^{\prime \prime }+(\gamma ^{\prime
})^{2}]g(r)=0,  \label{14}
\end{equation}%
which may be recast in the following operatorial form
\begin{equation}
\Bigg(\frac{d}{dr}+\gamma ^{\prime }\Bigg)\Bigg(-\frac{d}{dr}+\gamma
^{\prime }\Bigg)g(r)=0.  \label{15}
\end{equation}%
Hence we have $g(r)\sim e^{\gamma (r)}$ and $\alpha _{0}$ ends up as a
constant, say $\tilde{\alpha}$, by Eq. (\ref{12}). Therefore, the
dimensional reduction of (\ref{2}) leads straightforwardly to
\begin{equation}
S=-\frac{1}{4}\int_{-\infty }^{+\infty }\tilde{\alpha}^{2}G(\bar{\phi}%
)dr\int d^{4}xf^{\mu \nu }f_{\mu \nu }.  \label{17}
\end{equation}%
\newline
Obviously, if the $G(\bar{\phi})$ function is constant the coupling constant
multiplying the usual four-dimensional Maxwell Lagrangian blows up,
rendering a delocalized gauge field, that is, the gauge field zero-mode
permeates the whole bulk.

\subsection{Setting up the $G(\bar{\protect\phi})$ function}

The idea that a neutral scalar field might be effectively coupled to a gauge
field dates back from the observations of the anomalous decay $\pi ^{0}$ $%
\rightarrow $ $2\gamma $ mediated by virtual fermions. Such an effective
coupling was found by Schwinger \cite{SCHWINGER} together with an effective
coupling term of a scalar neutral field to the electromagnetic field. The
latter effective coupling would describe the decay of a stationary meson
into two parallel polarized photons mediated by a virtual proton-antiproton
pair, namely (see Eq. (5.6) of \cite{SCHWINGER}) $\mathcal{L}=(e^{2}/12\pi
)(g/M)\phi F^{\mu \nu }F_{\mu \nu }$. In trying to follow this clue it is
important to stress that the simple replacement $G(\bar{\phi})\varpropto
\bar{\phi}(r)$ is not satisfactory to our problem, since $\bar{\phi}(r)$ is
not normalizable in the entire domain of the extra dimension. This is a
peculiar feature of domain wall solutions, as the one given by Eq. (\ref{9}%
), whatever the nonlinear model one uses to describe thick branes.

Friedberg and Lee proposed a phenomenological model \cite{LEE} to explain
nonperturbative effects of QCD at low energies. In that model, hadrons are
nontopological solitons of a nonlinear field theory potential involving a
phenomenological scalar field, $\sigma $, which couples to the quarks by
means of a Yukawa coupling and to the gluons by means of a dielectric
function, namely $\mathcal{L}=(-1/4)\kappa (\sigma )F_{c}^{\mu \nu }F_{\mu
\nu }^{c}$. Without going into details, we just recall that in the
Friedberg-Lee model the functional dependence of $\kappa (\sigma )$ on $%
\sigma $ is not crucial, but it has to satisfy some conditions such that the
QCD vacuum works as dia-electric medium for the chromo electric field and an
anti-diamagnetic medium for the chromo magnetic field, in close analogy to
the Meissner effect in superconductors. Those conditions are $\kappa (0)=1$,
$\kappa (\bar{\sigma})=0$ and $d\kappa (\bar{\sigma})/d\sigma =0$, where $%
\bar{\sigma}$ is the expectation value of the scalar field on the QCD
vacuum. Such conditions might be suit to the $G(\bar{\phi})$ function. Here
we set $G(\bar{\phi})=1$ on the core of the brane, and $\tilde{\alpha}$ can
be conveniently chosen such that $\int_{-\infty }^{+\infty }\tilde{\alpha}%
^{2}G(\bar{\phi})dr=1$. The other two conditions over $G(\bar{\phi})$ are $G(%
\bar{\phi})\rightarrow 0$ and $dG(\bar{\phi})/d\bar{\phi}\rightarrow 0~$%
asymptotically ($r\rightarrow \pm \infty $), that is, when $\bar{\phi}%
(r\rightarrow \pm \infty )$ goes to the two respective neighbors minima of
the potential $V(\phi )$.

We have found that some functionals satisfy those conditions. At this point
we would like to recall that the warp factor itself, which keeps a
connection to $\bar{\phi}(r)$, plays the role of a smearing out weight
function to localize gravitons on branes \cite{GABA}, and it would also
satisfies the above conditions imposed over $G(\bar{\phi})$. Nevertheless,
since we want to localize gauge field on branes embedded in flat space-time
too, as in the Rubakov-Shaposhnikov scenario \cite{RUBA1}, we keep looking
for a functional of $\bar{\phi}(r)$. Particularly, in flat space-time Eq. (%
\ref{13}) reduces to%
\begin{equation}
-g_{n}^{^{\prime \prime }}(r)+\big[\gamma ^{\prime \prime }+(\gamma ^{\prime
})^{2}\big]g_{n}(r)=m_{n}^{2}g_{n}(r),  \label{18}
\end{equation}%
where $\gamma ^{\prime }=G^{\prime }/2G$ is the quantum mechanics
superpotential. Such equation is very similar to the equation for the
excitations of the brane (branons) around the domain wall solution. In this
last case the quantum mechanics superpotential is given by \cite{CVH}%
\begin{equation}
\gamma ^{\prime }=W_{\phi \phi }(\bar{\phi}(r)).  \label{QQUER}
\end{equation}%
Furthermore, Eqs. (\ref{18})-(\ref{QQUER}) also appear in the case of
fermion fields localization on branes in flat space-time when the coupling
of fermions to the scalar field is inspired on supersymmetry, that is, $%
W_{\phi \phi }\overline{\Psi }\Psi $. By keeping such recurrence also in the
case of localization of gauge fields on branes, we set%
\begin{equation}
\gamma ^{\prime }=G^{\prime }/2G=\kappa W_{\phi \phi }(\bar{\phi}(r)),
\end{equation}%
being $\kappa $ a positive constant, which leads to%
\begin{equation}
G(\bar{\phi}(r))\varpropto W_{\phi }^{2\kappa }(\bar{\phi}(r)).  \label{19}
\end{equation}

Now we are in position to set up the smearing out functions for both, flat
and warped space-time. In Ref. \cite{RUBA1} one has $V(\phi )=W_{\phi
}^{2}/2=(\lambda /4)(\phi ^{2}-m^{2}/\lambda )^{2}$ and $\bar{\phi}(r)=(m/%
\sqrt{\lambda })\tanh (mr/\sqrt{2})$. Therefore one obtains%
\begin{equation}
G(\bar{\phi}(r))=\text{sech}^{4\kappa }(mr/\sqrt{2}).  \label{FILO}
\end{equation}%
Eq. (\ref{FILO}) is the appropriated $G(\bar{\phi})$ function to the flat
space, according to our analogy and to the conditions imposed over $G(\bar{%
\phi}(r))$. Such a superpotential, however, is not adequate for brane worlds
scenario in warped space-time, because it implies into a unbound from below
potential $V(\phi )$ as given by Eq. (\ref{6a}). Nevertheless, by taking $%
W(\phi )$ as in Eq. (\ref{7}) together with the domain wall solution (\ref{9}%
), one finds%
\begin{equation}
G(\bar{\phi}(r))=\mathrm{sech}^{2\kappa }(2cr).  \label{QUILO}
\end{equation}

Both these smearing out functions, (\ref{FILO}) and (\ref{QUILO}), are sharp
on the core of the brane and exhibit a narrow bell shape profile, in such a
way that they are normalizable in the entire domain of the extra coordinate.

\section{Localizing the Kalb-Ramond field}

We start with the K-R lagrangian suitable modified by the multiplication of
the smearing out function

\begin{equation}
S=-\frac{1}{12}\int \sqrt{g}G(\bar{\phi})H_{MNL}H^{MNL},  \label{kr2}
\end{equation}
where
\begin{equation}
H_{MNL}= \partial_{M}B_{NL} + \partial_{N}B_{LM} + \partial_{L}B_{MN},
\label{krsl1}
\end{equation}
is the field strength for the K-R field.

The equation of motion for the field $B_{MN}$ is given by
\begin{equation}
\partial _{Q}(\sqrt{g}G(\bar{\phi})g^{MQ}g^{NR}g^{LS}H_{MNL})=0,
\label{krsl2}
\end{equation}%
which with the aid of Eq. (\ref{4}) can be expressed as
\begin{equation}
e^{2A}G(\bar{\phi})\partial _{\mu }H^{\mu \gamma \theta }-\partial _{y}(G(%
\bar{\phi})H^{y\gamma \theta })=0.  \label{krsl3}
\end{equation}%
With the gauge choice $B^{\mu r}=0$, $\partial _{\mu }B^{\mu \nu }=0$ and
decomposing the field as $B^{\gamma \theta }=\sum_{i=1}^{n}h^{\gamma \theta
}(x)U_{n}(r)$ we have%
\begin{equation}
m_{n}^{2}U_{n}(r)+e^{2A}\Big[U_{n}^{\prime \prime }(r)+\frac{G^{^{\prime }}(%
\bar{\phi})}{G(\bar{\phi})}U_{n}^{\prime }(r)\Big]=0.  \label{krsl4}
\end{equation}

Just as in the gauge field case, in order to set a typical quantum
mechanical problem, it is convenient to perform the following
transformation:
\begin{equation}
U_{n}(r)=e^{-\omega (r)}h_{n}(r).  \label{kr12}
\end{equation}%
Now, by means of the identification
\begin{equation}
\omega ^{\prime }=G(\bar{\phi})^{\prime }/2G(\bar{\phi}),  \label{krsl5}
\end{equation}%
we obtain a Schrödinger-like equation%
\begin{equation}
-h_{n}^{\prime \prime }(r)+(\omega ^{\prime \prime }+\omega ^{\prime
}{}^{2})h_{n}(r)=m_{n}^{2}e^{-2A}h_{n}(r).  \label{krsl6}
\end{equation}%
For the massless zero mode $(h_{0}\equiv h)$ we have simply
\begin{equation}
-h^{\prime \prime }(r)+(\omega ^{\prime \prime }+\omega ^{\prime
}{}^{2})h(r)=0,  \label{krsl7}
\end{equation}%
which may be rewritten in the operatorial form
\begin{equation}
\Bigg(\frac{d}{dr}+\omega ^{\prime }\Bigg)\Bigg(-\frac{d}{dr}+\omega
^{\prime }\Bigg)h(r)=0.  \label{kr15}
\end{equation}%
Hence we have $h(r)\sim e^{\omega (r)}$ and by means of Eq. (\ref{kr12}), $%
U_{0}(r)$ ends up as a constant, say $\alpha $. Therefore, the dimensional
reduction of (\ref{kr2}) leads directly to
\begin{equation}
S=-\frac{1}{12}\int_{-\infty }^{+\infty }dr\alpha ^{2}e^{-2A}G(\bar{\phi}%
)\int d^{4}xh_{\alpha \beta \gamma }h^{\alpha \beta \gamma }.  \label{krimp}
\end{equation}

In order to reproduce an asymptotic AdS bulk, the warp factor $e^{2A}$ have
a gaussian-like shape peaked at the core of the brane, then $%
e^{-2A(r)}\rightarrow \infty $ as $r\rightarrow \pm \infty $, for all models
used to describe thick branes, and that is the reason for not a having a
localized zero mode. Hence, if $G(\bar{\phi})$ is again a convenient
smearing out function of $r$, it would be possible to localize the K-R zero
mode on the brane. Such a smearing out function would also work for flat
space ($e^{-2A(r)}=1$), rendering a localized tensorial field.

\subsection{Identifying the smearing out function}

The first clue we shall follow in order to set a suitable $G$ function is
the fact that by means of Eqs. (\ref{krsl5}) and (\ref{krsl7}), a given $G$
modify the quantum mechanics potential acting on the modes for the K-R field.

Following this reasoning and the recurrence mentioned in the previous
section, we shall identify the $G$ as in Eq. (\ref{19}) and check what would
be the constraints over $\kappa $ which make $\int_{-\infty }^{+\infty
}dr\alpha ^{2}e^{-2A(r)}G(\bar{\phi})$ convergent. We have noted that the
conditions over $\kappa $ is very dependent on the model we are using to
describe thick branes. We illustrate that by resorting to the same models we
have used in the previous section.

For the case $W_{\phi }^{2}=(\lambda /2)(\phi ^{2}-m^{2}/\lambda )^{2}~$we
have $\bar{\phi}(r)=(m/\sqrt{\lambda })\tanh (mr/\sqrt{2})$ and
\begin{eqnarray}
e^{-2A}W_{\phi }^{2k} &\varpropto &\left. \mathrm{sech}^{4\kappa +\frac{8}{9}%
\frac{m^{2}}{\lambda }}\left( \frac{mr}{\sqrt{2}}\right) \right.  \notag \\
&\times &\left. e^{\frac{2m^{2}}{9\lambda }\tanh ^{2}(\frac{mr}{\sqrt{2}}%
)}.\right.  \label{essa}
\end{eqnarray}%
Hence, upon integration over the extra dimension the Eq. (\ref{essa}) is
convergent for $\kappa \geq (-2/9)m^{2}/\lambda $. Since $\kappa $ is
positive, it is always convergent in this case and the localization of the
zero mode for the K-R field is accomplished without any restriction.

Now, keeping in mind the background given in Ref. (\cite{MG}) it is easy to
see that the integration along the extra dimension
\begin{equation}
\int_{-\infty }^{+\infty }dr\,e^{-2A}W_{\phi }^{2k}\varpropto \int_{-\infty
}^{+\infty }dr\,\mathrm{sech}^{2(\kappa -b)}(2cr)  \label{krult7}
\end{equation}%
is convergent whenever $\kappa >b$. Therefore, differently from the (\ref%
{essa}) case, here we have a nontrivial constraint over $\kappa $ which must
be fulfilled in order to localize the zero mode for the K-R field. One can
note that there will be no restriction over $\kappa $ if one works in flat
geometry, since there is no warp-factor.

\section{Final Remarks and Outlook}

We have proposed a mechanism that leads to gauge field zero mode
localization on thick branes, by means of an effective model obtained via
the introduction of a smearing out $G(\phi )$ function in the gauge field
Lagrangian. $G(\phi )$ is a functional of the classical scalar-gravitational
field equations solution which originates the brane in a warped space-time,
but the procedure can be applied to the case of flat space-time as well.

In order to set up a physically motivated $G(\phi )$ function, we rely on
the Friedberg-Lee phenomenological model proposed to explain nonperturbative
effects of QCD at low energies. This model involves a scalar field coupling
(via a Yukawa term) to the quarks and also coupling to the gluons by means
of a dielectric function. Translating to our problem, the analog $G(\phi )$
function plays the role of a smearing out dieletric function.

The case of flat geometry is more manageable in the determination of $G(\phi
)$ and we are guided by the problem of matter fields localization on branes.
By leading this recurrence a little further we were able to identify the
smearing out function as $G(\bar{\phi}(r))\varpropto W_{\phi }^{2\kappa }(%
\bar{\phi}(r))$, where $\kappa $ is a positive constant. Such functional
form to the $G$ function is suitable for gauge field localization for both,
flat and warped geometries.

It is worthwhile to mention that this simple effective model, based upon
phenomenological quantum field theory scenarios, might also be applied to
the localization of Yang-Mills fields on branes. In fact, we believe that,
giving the root of the Friedberg-Lee model itself, the extension of this
analogy to the non-Abelian gauge fields localization follows
straightforwardly.

The very same procedure is adopted to get the localization of Kalb-Ramond
fields on a thick brane. In that case we have found that more restrictive
conditions over $\kappa $ are necessary in order to accomplish the
localization and also that such restrictions depend on the model one has in
hands to describe thick branes.

We also have found that there is a mapping from the quantum mechanics
resulting from our approach namely, equations (\ref{13}) and (\ref{krsl4}),
into the quantum mechanics for the localized and resonant modes for the
vector and tensor gauge fields in dilatonic branes, which were carried out
in references \cite{CTA}, \cite{TCA} and \ \cite{CCT}. In the latter, the
quantum mechanics potentials for the excitations associated to the vector
and tensor gauge fields depends on the warp factor and on $A^{\prime }(r)$, $%
\pi ^{\prime }(r)\varpropto A^{\prime }(r)$ and $B^{\prime }(r)\varpropto
A^{\prime }(r)$, where $\pi (r)$ is the dilaton field and $e^{2B(r)}$ is an
extra warp factor from the metric used in the models for dilatonic thick
branes. We have noted that the dependence on those terms is such that their
resulting quantum mechanics potential is proportional to the quantum
mechanics potential found in our approach, provided that the same nonlinear
field theory model is used to describe the thick branes in both cases. Such
a mathematical mapping is much clear when one deals with the model defined
by the superpotential\ (\ref{7}), because in this case the term $G(\bar{\phi}%
)^{\prime }/G(\bar{\phi})$ is proportional to $A^{\prime }(r)$. Such a
relation can be used to develop a straightforward analysis of the resonant
modes for the vector and tensor gauge fields in our case, by resorting to
the results found in \cite{CTA}, \cite{TCA} and \ \cite{CCT}. We think that
our results, concerning resonant modes, will not differ appreciably from
theirs.

\section*{Acknowledgments}

AERC thanks to CAPES for financial support. MBH thanks to CNPq for partial
support.

\end{document}